\documentstyle[12pt]{article}
\newcommand{\ba}[1]{\begin{eqnarray} \label{#1}}
\newcommand{\ea}{\end{eqnarray}}
\newcommand{\nn}{\nonumber}
\def\half {\frac{1}{2}}
\def\WT{\widetilde}
\def\ifmath#1{\relax\ifmmode #1\else $#1$\fi}
\def\ls#1{\ifmath{_{\lower1.5pt\hbox{$\scriptstyle #1$}}}}
\begin{document}
\begin{center}
{\Large \bf 
	SUSY spectrum constraints on direct dark matter detection
\par }

\bigskip

{\large V.A.~Bednyakov\footnote{Laboratory of Nuclear Problems,
	Joint Institute for Nuclear Research,
        Moscow region, 141980 Dubna, Russia. E-mail: bedny@nusun.jinr.ru} 
	and H.V.~Klapdor-Kleingrothaus }
\bigskip

        {\it
        Max-Planck-Institut f\"{u}r Kernphysik, \\
        Postfach 103980, D-69029, Heidelberg, Germany
        }
\end{center}

\begin{abstract}
	We perform an investigation of the MSSM parameter space 
	at the Fermi scale taking into account
	available accelerator, non-accelerator and cosmological constraints.
	Extra assumptions about upper bounds
	for some of the SUSY particles are also imposed.
	We show that a non-observation of the SUSY dark matter
	candidates with a high-accuracy dark matter detector, 
	such as $^{73}$Ge, under above-mentioned assumptions can
	exclude large domains of the MSSM parameter space 
	and, for instance, can make 
	especially desirable 	
	collider search for light SUSY charged Higgs boson.
\end{abstract}

	As already well-known 
\cite{DressNojiriRate,KKSbook,JuKaGr}, a direct dark matter search for 
	neutralinos, lightest SUSY particles (LSP), is complimentary
	to high energy searches for SUSY with colliders.
	The direct dark matter search is able to give 
	information which is not available from collider physics.

	It has been found, that future-possible
	SUSY spectrum restrictions from colliders,
	especially such as upper mass bounds for some of the SUSY
	particles, can strongly enforce the importance of 
	the direct dark matter detection experiments.
	The main point is that this hypothetical information  
	leads to some important restrictions 
	of the expected event rates for direct dark matter detection.
	In some cases the restriction is a clear lower limit for this rate.
	Another obvious example of such restriction is 
	an upper bound for the LSP, which appears only due to the fact
	that all sfermions and gauginos are not lighter than the LSP.
	Thus, these high-energy bounds bring in restrictions on
	the parameter space which can result in
	a lower limit for the expected rate.
	Such effect is obtained mostly when the sfermion masses 
	are restricted, but the effect appeared quite strongly especially 
	in the Higgs sector of the MSSM
\cite{haka}.

	The question --- 
	to what extent the accelerator searches are able to 
	affect future prospects for direct dark matter search provided 
	the dark matter particle candidate is LSP --- 
	was investigated comprehensively 
	in the minimal supergravity SUSY models 	
\cite{9810279,9706509,9801246,9710238}.
	In this paper we explore the MSSM parameter space at the weak
	scale in the most phenomenological way, relaxing completely
	any constraints following from the unification assumption.
	On the  other side we hold all restrictions from the age of
	the Universe, accelerator SUSY searches, rare FCNC 
	$b\to s\gamma$ decay, etc
\cite{9801246,9710238,roskane}.

	The MSSM parameter space in our approach is determined 
	by entries of the mass matrices of neutralinos, charginos, 
	Higgs bosons, sleptons and squarks.
	To specify the parameter space we give all relevant
	mass matrices below.
	The one-generation squark and slepton mass matrices  
	(using the notation of the third family) are given by
\cite{haka}
\ba{Squarks} \nn
        M^2_{\tilde t} &=& \left[ \matrix{%
        M^2_{\widetilde Q}+ m^2_t+ m^2_Z(\half-\frac23 s^2_W)\cos2\beta
        & m_t(A_t-\mu\cot\beta) \cr
        m_t(A_t-\mu\cot\beta)
        &M^2_{\widetilde U}+ m^2_t+ m^2_Z \frac23 s^2_W\cos2\beta \cr}
      \right], \\
\nn
        M^2_{\tilde b} &=& \left[ \matrix{%
    M^2_{\widetilde Q}+ m^2_b- m^2_Z(\half-\frac13 s^2_W)\cos2\beta
      & m_b(A_b-\mu\tan\beta) \cr
    m_b(A_b-\mu\tan\beta)
      &M^2_{\widetilde D}+ m^2_b - m^2_Z \frac13 s^2_W\cos2\beta \cr }
      \right],
\\ \nn
  M^2_{\tilde\nu} &=& M^2_{\widetilde L} + \half m^2_Z\cos2\beta, \\
\nn 
  M^2_{\tilde \tau}  &=& \left[ 
                        \begin{array}{cc}
    M^2_{\widetilde L}+ m^2_\tau- m^2_Z(\half- s^2_W)\cos2\beta
      & m_\tau(A_\tau-\mu\tan\beta) \\
    m_\tau(A_\tau-\mu\tan\beta)
      &M^2_{\widetilde E}+ m^2_\tau- m^2_Z s^2_W\cos2\beta 
                        \end{array}
      \right]
\ea
	where $s^2_W\equiv\sin^2\theta_W$ and
	$\tan\beta \equiv \langle {H^0_2} \rangle / \langle {H^0_1}\rangle$.
	In the $\WT W^+$--$\WT H^+$ basis, the chargino mass matrix is 
$$
  X = \pmatrix{ M_2 &\sqrt 2 m\ls W \sin\beta \cr
       \sqrt 2 m\ls W \cos\beta &\mu \cr}.  
$$
	Two unitary $2\times 2$ matrices $U$ and $V$ are required
	to diagonalize the chargino mass-squared matrix
$
{\cal M}_{\WT\chi^+}^2=VX^\dagger XV^{-1}=U^\ast XX^\dagger
(U^\ast)^{-1}.
$
	The two mass eigenstates are denoted by $\widetilde\chi^+_1$
	and $\widetilde\chi^+_2$.
	In the $\WT B$--$\WT W^3$--$\WT H^0_1$--$\WT H^0_2$ basis, the
	neutralino Majorana mass matrix is
$$
  Y = \pmatrix{%
    M_1 &0  &-m_Zc_\beta s_W &m_Zs_\beta s_W \cr
    0   &M_2 &m_Z c_\beta c\ls W &-m_Z s_\beta c\ls W \cr
    -m_Z c_\beta s\ls W &m_Z c_\beta c\ls W &0 &-\mu \cr
    m_Z s_\beta s\ls W &-m_Z s_\beta c\ls W &-\mu &0 \cr }\,,
$$
	where $s_\beta = \sin\beta$, $c_\beta=\cos\beta$, etc.  
	A $4\times 4$ unitary matrix $Z$ is required to diagonalize 
	the neutralino mass matrix
	${\cal M}_{\WT\chi^0}=Z^\ast YZ^{-1} $
	where the diagonal elements of ${\cal M}_{\WT\chi^0}$ can 
	be either positive or negative.
	The CP-even Higgs mass matrix has the form
\cite{Higgses}
\begin{eqnarray*}
	\half
        \left(\frac{\partial^2 V}{\partial \psi_i \partial \psi_j}
        \right)_{v_1,v_2}
         &\equiv&
         \left(\begin{array}{cc}
               H_{11} & H_{12}\\  
               H_{12} & H_{22} 
          \end{array}\right)
	=
	\half \left(\begin{array}{cc}
                 \tan\beta & -1 \\   -1 & \cot\beta 
         \end{array}\right) 
                  M_A^2\sin 2\beta                
\\ \nn
      &+&\half \left(\begin{array}{cc}
                  \cot\beta & -1 \\   -1 & \tan\beta 
           \end{array}\right) m_Z^2\sin 2\beta 
       +\omega
         \left(\begin{array}{cc}
               \Delta_{11} & \Delta_{12}\\  
               \Delta_{12} & \Delta_{22} 
          \end{array}\right),
\end{eqnarray*}
	where $\omega = \frac{3g_2^2}{16\pi^2m_W^2}$ and: 
$$ H^{}_{11} = \frac{\sin 2\beta}{2}
        ( \frac{m^2_Z}{\tan\beta} + M^2_A\tan\beta )
                + \omega\Delta_{11},
$$
$$ H^{}_{22} = \frac{\sin 2\beta}{2}
       ( m^2_Z \tan\beta + \frac{M^2_A}{\tan\beta} )
                + \omega\Delta_{22},
$$
$$ H^{}_{12} = H^2_{21} = - \frac{\sin 2\beta}{2}
        ( m^2_Z + M^2_A )
                + \omega\Delta_{12}.
$$
\ba{d11}
\nn
 \Delta_{11} &=& \frac{m^4_b}{c^2_\beta}
 (  \ln{\frac{m^2_{\tilde{b}_1} m^2_{\tilde{b}_2}} {m^4_b}}
  + \frac{2 {A}_b( {A}_b-\mu \tan\beta)}
         {m^2_{\tilde{b}_1}-m^2_{\tilde{b}_2}}
   \ln{ \frac{m^2_{\tilde{b}_1}} {m^2_{\tilde{b}_2}} }  ) \\
\nn   &+& \frac{m^4_b}{c^2_\beta}
( \frac{ {A}_b( {A}_b-\mu \tan\beta)}
      {m^2_{\tilde{b}_1}-m^2_{\tilde{b}_2}})^2
 g_{sf}(m^2_{\tilde{b}_1}, m^2_{\tilde{b}_2} ) \\
\nn   &+& \frac{m^4_t}{s^2_\beta}
 ( \frac{ \mu( {A}_t-\frac{\mu}{\tan\beta})} 
        {m^2_{\tilde{t}_1}-m^2_{\tilde{t}_2}})^2
 g_{sf}(m^2_{\tilde{t}_1}, m^2_{\tilde{t}_2} ).
\\ \nn
 \Delta_{22} &=& \frac{m^4_t}{s^2_\beta}
 ( \ln{ \frac{ m^2_{\tilde{t}_1} m^2_{\tilde{t}_2} }
             { m^4_t                               }
      }
     + \frac{ 2 {A}_t( {A}_t-\frac{\mu}{\tan\beta}) }
             { m^2_{\tilde{t}_1}-m^2_{\tilde{t}_2}         }
             \ln{ \frac{ m^2_{\tilde{t}_1} }
                       { m^2_{\tilde{t}_2} }
                }
 )      \\
\nn      &+&  \frac{m^4_t}{s^2_\beta}
    ( \frac{ {A}_t ( {A}_t-\frac{\mu}{\tan\beta} ) }
          { m^2_{\tilde{t}_1}-m^2_{\tilde{t}_2}          }
   )^2  g_{sf}( m^2_{\tilde{t}_1},m^2_{\tilde{t}_2} ) \\
\nn      &+&  \frac{m^4_b}{c^2_\beta}
   ( \frac{ \mu ( {A}_b-\mu\tan\beta )       }
          {m^2_{\tilde{b}_1} - m^2_{\tilde{b}_2} }
   )^2 g_{sf}( m^2_{\tilde{b}_1},m^2_{\tilde{b}_2} ).
\\ \nn
  \Delta_{12} &=& \Delta_{21} = \frac{m^4_t}{s^2_\beta}
     \frac{\mu ( {A}_t-\frac{\mu}{\tan\beta}) }
          { m^2_{\tilde{t}_1}-m^2_{\tilde{t}_2} }
  (  \ln{ \frac{ m^2_{\tilde{t}_1} }
               { m^2_{\tilde{t}_2} }
        }
      + \frac{  {A}_t( {A}_t-\frac{\mu}{\tan\beta})}
             {  m^2_{\tilde{t}_1} - m^2_{\tilde{t}_2}    }
      g_{sf}(  m^2_{\tilde{t}_1}, m^2_{\tilde{t}_2}   )
  )\\
\nn    &+& \frac{m^4_b}{c^2_\beta}
     \frac{\mu ( {A}_b- \mu\tan\beta) }
          { m^2_{\tilde{b}_1}-m^2_{\tilde{b}_2} }
  (  \ln{ \frac{ m^2_{\tilde{b}_1} }
               { m^2_{\tilde{b}_2} }
        }
      + \frac{ {A}_b( {A}_b-\mu\tan\beta) }
             {  m^2_{\tilde{b}_1} - m^2_{\tilde{b}_2}    }
      g_{sf}(  m^2_{\tilde{b}_1}, m^2_{\tilde{b}_2}   )
  ).
\ea
	Here $c^2_\beta = \cos^2\!\beta$, $s^2_\beta = \sin^2\!\beta$ 
	and
$ g_{sf}(m_1^2,m_2^2) =  2 - \frac{m_1^2+m_2^2}{m_1^2-m_2^2}
                       \ln{ \frac{m_1^2}{m_2^2} }.
$
	Neutral CP-even Higgs eigenvalues are	
$$
m_{H,h}^2 = \half \Bigl\{ H_{11}+H_{22}
                \pm \sqrt{(H_{11}+H_{22})^2 - 4(H_{11}H_{22}-H_{12}^2)} 
                  \Bigr\},
$$
	The mixing angle $\alpha_H^{}$ is obtained from
$$
   \sin 2\alpha_H^{}= \frac{2H^2_{12}}{m^2_{H^0_1} - m^2_{H^0_2}},
\qquad
   \cos 2\alpha_H^{}= \frac{H^2_{11}-H^2_{22}}{m^2_{H^0_1} - m^2_{H^0_2}}.
$$ 
	The mass of the charged Higgs boson is given by
  	$m^2_{\rm CH} = m^2_W + M^2_A   + \omega\Delta_{\rm ch}$.

	Therefore free parameters are: $\tan\beta$ is the ratio
	of neutral Higgs boson vacuum expectation values, $\mu$ is 
	the bilinear Higgs parameter of the superpotential,
	$M_1$, $M_2$ are soft gaugino masses, $M_A$ is the CP-odd 
	Higgs mass, $m^2_{\WT Q}$, $m^2_{\WT U}$, $m^2_{\WT D}$ are 
	squared squark mass parameters for the 1st and 2nd generation,        
	$m^2_{\WT L}$, $m^2_{\WT E}$ are squared 
	slepton mass parameters for the 1st and 2nd generation, 
	$m^2_{\WT Q_3}$, $m^2_{\WT T}$, $m^2_{\WT B}$ are squared
	3rd generation squark mass parameters, $m^2_{\WT L_3}$, 
	$m^2_{\WT \tau}$ are squared 3rd generation slepton mass 
	parameters and $A_t$, $A_b$, $A_\tau$ are soft trilinear 
	couplings for the 3rd generation.
	With these parameters one completely determines the SUSY spectrum
	and the MSSM coupling constants at the Fermi scale.
	
        A dark matter (DM) event is elastic scattering 
	of a relic DM neutralino from a target nucleus producing a nuclear 
	recoil which can be detected by a suitable detector
\cite{JuKaGr}.
	The differential event rate in respect to the recoil 
	energy is the subject of experimental measurements.
	The rate depends on the distribution of
        the DM neutralinos in the solar vicinity and
        the cross section of neutralino-nucleus elastic scattering.

\begin{figure}[t!] 
\begin{picture}(100,145)
\put(20,-5){\includegraphics{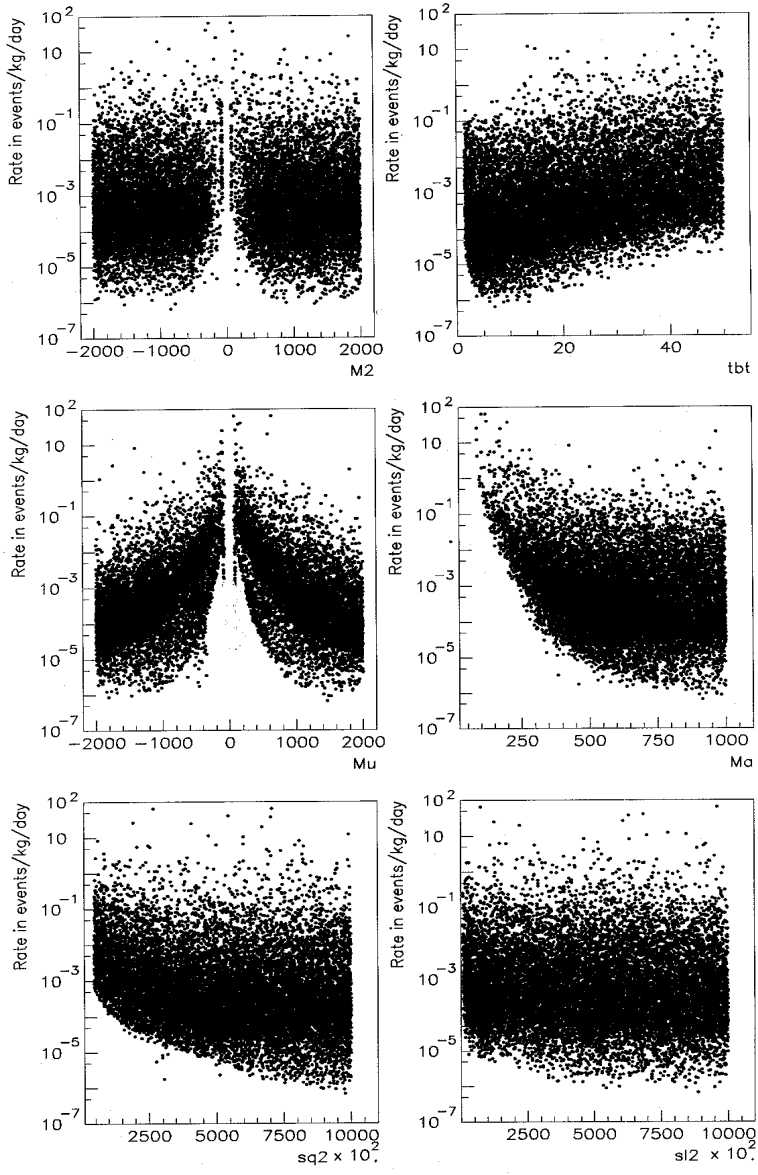}}
\end{picture}
\caption{Total event rate in $^{73}$Ge versus MSSM parameters 
	 $M^{}_2$, $\tan\beta$, $\mu$, $M^{}_A$, $m^2_{Q_{1,2}}$ 
	(labeled as "sq2") and $m^2_{L}$ (as "sl2").
}
\label{Rate-scan}
\end{figure}

\begin{figure}[t!] 
\begin{picture}(100,145)
\put(20,-5){\includegraphics{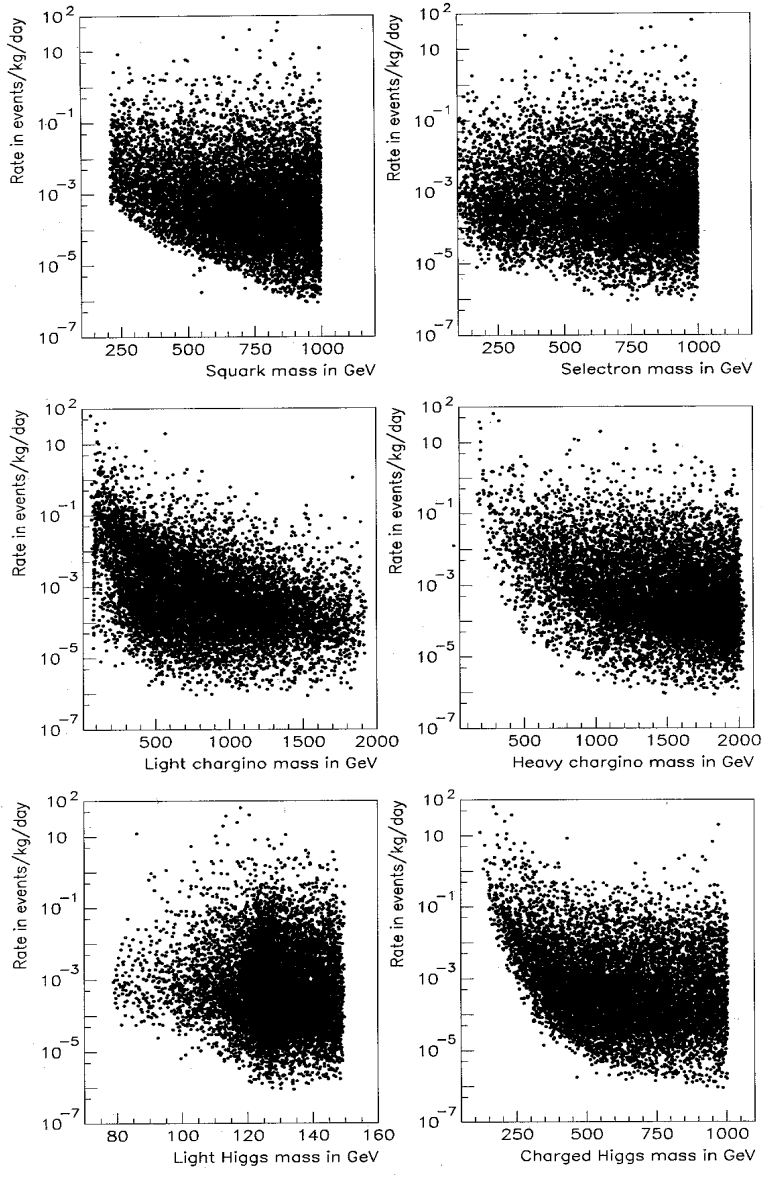}}
\end{picture}
\caption{Total event rate in $^{73}$Ge versus masses of squarks, 
	sleptons, light chargino (practically the same scatter plot 
	one obtains for next-to-lsp neutralino), heavy chargino, 
	light CP-even neutral Higgs boson and charged Higgs boson 
	(the same is for heavy Higgs boson).
}
\label{Rate-spectrum}
\end{figure}

	In our analysis we use the so-called total event rate $R$ 
	which is integrated over recoil energies, and which is useful 
	for searching for domains with extreme rates.
	We follow our paper
\cite{Superlight}, where on can find all relevant formulas
	and astrophysical parameters.
	To calculate the event rate we use for the relic neutralino 
	mass density and for the escape neutralino velocity commonly 
	accepted values 0.3~GeV$/$cm$^3$ and 600 km$/$s, respectively.
	Their experimental variations can slightly change $R$ but 
	leave the dependence of $R$ on the MSSM parameters 
(Fig.~\ref{Rate-scan} and 
\ref{Rate-spectrum})  unaffected.
	To compare our results with sensitivities of different
	dark matter experiments we calculate also the total cross 
	section for relic neutralino elastic scattering on the nucleon.

   	The present lifetime of the Universe implies an upper limit 
	on the expansion rate and correspondingly on the total 
	relic abundance.
	One finds 
\cite{kolb} that  the contribution of each relic particle species 
	$\chi$  has to obey $\Omega_\chi h^2_0<1$, where $h_0$ is the 
	Hubble constant and  the relic density parameter  
	$\Omega_\chi = \rho_\chi/\rho_c$ is the ratio of the relic 
	neutralino mass  density $\rho_\chi$ to  the critical one
   	$\rho_c = 1.88\cdot 10^{-29}$h$^2_0$g$\cdot$cm$^{-3}$.
 	Assuming that the neutralinos form a dominant part of
    	the dark matter in the Universe one obtains 
	a lower limit on the neutralino relic density.
	We restrict our analysis with the cosmological constraint 
	$0.025 < \Omega_\chi h^2_0<1$
\cite{DressNojiriRate,9801246,Superlight}.
   	We calculate $\Omega_{\chi} h^2_0$  following the standard
   	procedure on the basis of the approximate formula
\cite{drno,Hagelin}. 
	We take into account all channels of the $\chi-\chi$ annihilation.
   	Since the neutralinos are mixtures of gauginos and
   	higgsinos, the annihilation can occur both, via
   	s-channel exchange of the $Z^0$ and Higgs bosons and
   	t-channel exchange of a scalar particle, like a selectron 
\cite{relic}.
   	This constrains the parameter space
\cite{roskane,drno,Bot}.

        Another stringent constraint is imposed by the 
	branching ratio of $b\to s\gamma$ decay, 
	measured by the CLEO collaboration to be 
	$1.0 \times 10^{-4} < {\rm B}(b\to s \gamma) < 4.2 \times 10^{-4}$.
        In the MSSM this flavor changing neutral current 
        process receives contributions from $H^\pm$--$t$,
        $\tilde{\chi}^\pm$--$\tilde{t}$ and $\tilde{g}$--$\tilde{q}$ loops
        in addition to the standard model \ \ $W$--$t$ loop.
	These also strongly restrict the parameter space
\cite{BerBorMasRi}.

	The masses of the supersymmetric particles are constrained 
	by the results from the high energy colliders
	LEP at CERN and Tevatron at Fermilab.
	This  imposes relevant constraints on 
	the parameter space of the MSSM.
	We use the following experimental restrictions 
	for the SUSY particle spectrum in the MSSM
\cite{PDG}:
$ M_{\tilde{\chi}^+_{2}} \geq 65$~GeV for the light chargino,
$ M_{\tilde{\chi}^+_{1}} \geq 99$~GeV for the heavy chargino,
$ M_{\tilde{\chi}^0_{1,2,3}} \geq 45, 76, 127$~GeV for non-LSP
					neutralinos, respectively;
$ M_{\tilde{\nu}}      \geq 43$~GeV for sneutrinos,
$ M_{\tilde{e}_R}      \geq 70$~GeV for selectrons,
$ M_{\tilde{q}}       \geq 210$~GeV  for squarks,
$ M_{\tilde{t}_1}      \geq 85$~GeV  for light top-squark,   
$ M_{H^0}              \geq 79$~GeV  for neutral Higgs bosons,
$ M_{\rm CH}           \geq 70$~GeV  for charged Higgs boson.

\enlargethispage{\baselineskip}
	In our numerical analysis the parameters of the MSSM are 
	randomly varied at the Fermi scale in the intervals given below
\begin{center}
\begin{tabular}{c@{~~$<$~~}c@{~~$<$~~}c}
  $-1000$~GeV  & $M_1$ &    1000~GeV   \\
  $-2000$~GeV  & $M_2$ &    2000~GeV   \\
        1      & $\tan\beta$      &      50   \\
  $-2000$~GeV  & $\mu$      &2000~GeV   \\
   $60$~GeV     & $M_A$  & 1000~GeV   \\
  10~GeV$^2$  & $m^2_{Q_{1,2}} $    & 1000000~GeV$^2$  \\
  10~GeV$^2$  & $m^2_{L} $    & 1000000~GeV$^2$  \\
  10~GeV$^2$  & $m^2_{Q_3} $    & 1000000~GeV$^2$  \\
  10~GeV$^2$  & $m^2_{L_3} $    & 1000000~GeV$^2$  \\
  $-2000$~GeV  & $A_t$       &    2000~GeV.
\end{tabular}
\end{center}
	For each parameter 
	some number, between 0 and 1, was defined by 
	the random generator from the CERN Library.
	With this number and 
	lower and upper bounds given above  
	the random value of the MSSM parameter was calculated 
	by means of linear function. 
	We stopped running when the lower border of the rate 
	$R$ stopped move down 
	and only density of points continued to increase. 
	In our main scan we have tested about $10^8$ models.
	and only about $10^5$ models passed all constraints.

	For simplicity, for other sfermion mass parameters we used 
	the relations $m^2_{\WT U_{1,2}} = m^2_{\WT D_{1,2}} 
	= m^2_{\WT Q_{1,2}}$, $m^2_{\WT E_{1,2}} = m^2_{L}$, 
	$m^2_{\WT T} = m^2_{\WT B} = m^2_{Q_3}$,  
	$m^2_{\WT E_{3}} = m^2_{L_3}$. 
	The parameters $A_b$ and $A_\tau$ are fixed to be zero.
	We consider the domain of the MSSM parameter space, 
	in which we perform our scans, as quite spread and natural. 
	Any extra expansion of it like, for example, using
	$-10$~TeV$< M_2< $~10~TeV, etc, of course, can be possible, 
	but should be considered as quite unnatural
	in the framework of the idea of SUSY.  

\begin{figure}[t!] 
\begin{picture}(100,140)
\put(12,-25){\includegraphics{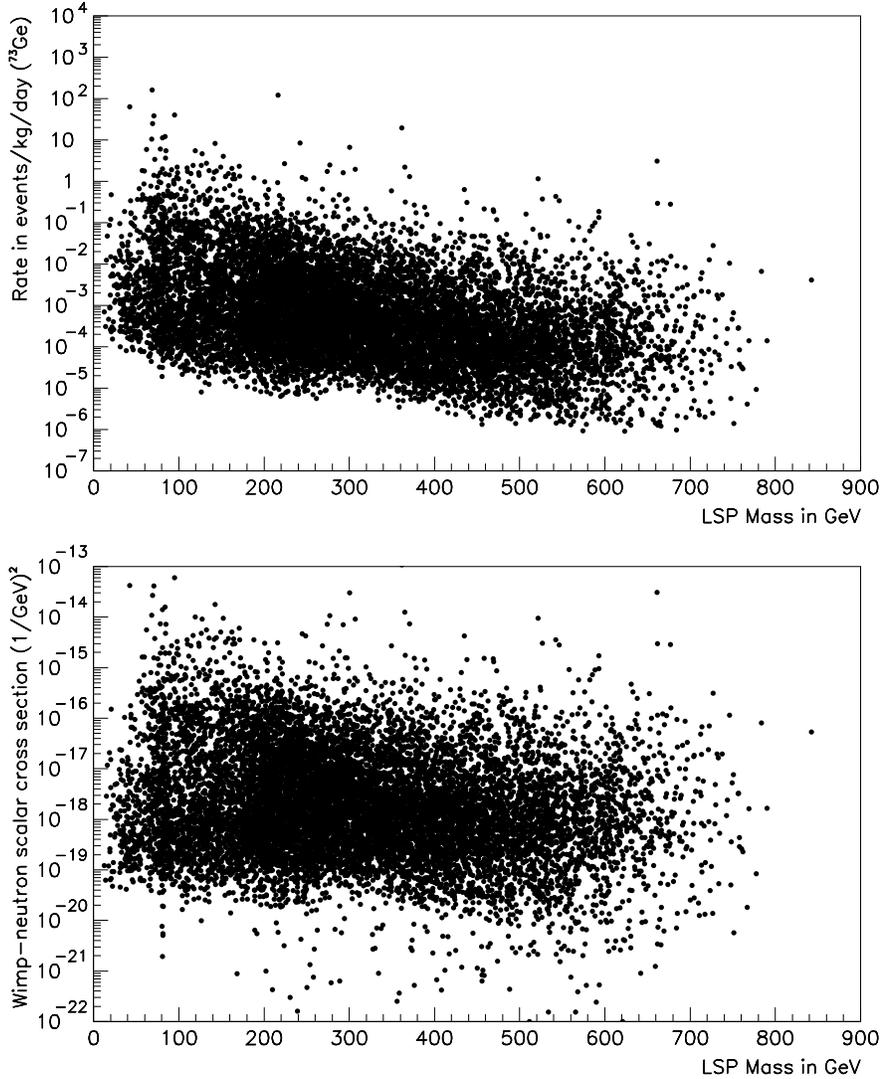}}
\end{picture}
\caption{Total event rate in $^{73}$Ge in events/day/kg (upper panel) 
	and cross section of scalar elastic scattering
	of LSP (WIMP) off neutron in GeV$^{-2}$ (lower panel) 
	versus mass of LSP (in GeV).}
\label{Rate-WIMP}
\bigskip
\end{figure}

    	Some of results of the main (without extra constraints) 
	scan are presented in
Figs.~\ref{Rate-scan}--\ref{Rate-WIMP} as scatter plots.
	The main feature we paying our attention to is the presence
	of a lower bound for the total event rate $R$. 
	An absolute minimum value of about $10^{-6}$ events/day/kg 
	in a $^{73}$Ge detector is obtained in the above-mentioned
	domain of the MSSM parameter space.
	There is a clear growth (up to one order of magnitude) 
	of the lower bound only with $\tan\beta$ 
(Fig.~\ref{Rate-scan}).
	In all other cases there is a decrease of the lower bound, 
	the decrease is most sharp with $|\mu|$, $M^{}_A$ 
	(about 5 orders of magnitude) and $M^2_{Q_{1,2}}$
(Fig.~\ref{Rate-scan}) and with squark mass $M^{}_{\WT q}$,  
	heavy chargino mass $M_{\tilde{\chi}^+_{1}}$
	and charged Higgs boson mass $M_{\rm CH}$ 
(Fig.~\ref{Rate-spectrum}).

Figure~\ref{Rate-WIMP} (upper panel)
	gives expectations for the total event rate $R$ obtained 
	by means of scanning the parameter space at the Fermi scale
	without any extra limitations on the SUSY particle spectrum.
 	There is a lower bound, which decreases with mass of the 
	LSP and reaches an absolute minimum of about $10^{-6}$ events/day/kg 
	in a $^{73}$Ge detector in the region of LSP masses 600--700 GeV.
	The phenomenologically allowed masses of the LSP are spread from 
	about 5 GeV till about 800 GeV.
	Considering both scatter plots in 
Fig.~\ref{Rate-WIMP} one also can conclude that, due to the 
	practical absence of a lower bound for the scalar cross 
	section of WIMP-neutron interaction, the lower bound for the 
	rate is mainly established by the spin-dependent interaction, 
	which in contrast to the scalar interaction is associated 
	with an about 3-order-of-magnitude larger lower bound for 
	WIMP-nucleon cross section.

\begin{figure}[th!] 
\begin{picture}(100,110)
\put(20,-2){\includegraphics{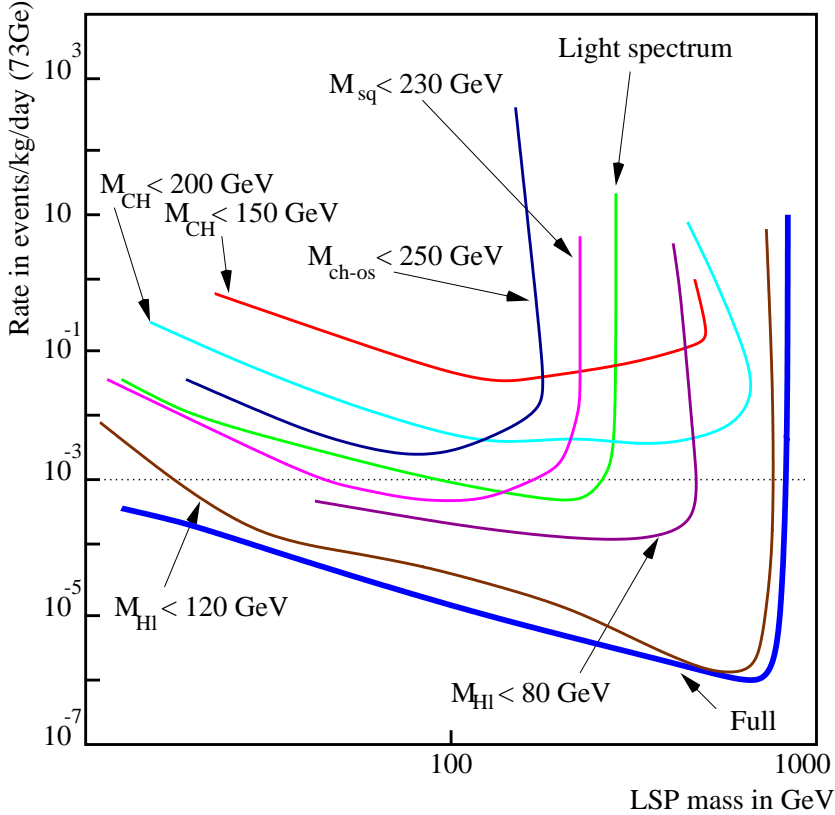}}
\end{picture}
\caption{Different lower bounds for the total  event rate in 
	$^{73}$Ge (events/day/kg)  versus mass of the LSP (GeV).
	Here M$^{}_{\rm sq,\ CH,\ Hl}$ denote masses of the squark,
	the charged Higgs boson and the light neutral CP-even Higgs boson 
	respectively.
	Heavy chargino mass is denoted as M$^{}_{\rm ch-os}$.
	"Full" corresponds to the lower bound obtained 
	from main (unconstrained) scan, and
	"Light spectrum" denotes the lower bound for $R$, which
	is obtained with all sfermion masses lighter then 
	about 300 GeV.
	The horizontal dotted line represents expected sensitivity
	for the direct dark matter detection with GENIUS.}
\label{LowBounds}
\end{figure}

	The existence of the lower bound for the event rate
	itself and the variation of the bound with the MSSM 
	parameters and masses of the SUSY particles 
	allow us to consider prospects to search for
	dark matter under special assumptions about 
	restricted values for the above-mentioned parameters and masses.
	To this end we have performed a number of extra scans
	taking into account extra limitations
	on single squark mass (M$^{}_{\rm sq}< 250,\ 230$~GeV), 
	light neutral CP-even Higgs boson mass 
	(M$^{}_{\rm Hl}< 80, \ 100, \ 120$~GeV), 
	charged Higgs boson mass (M$^{}_{\rm CH}<150, \ 200$~GeV) 
	and heavy chargino mass (M$^{}_{\rm ch-os}< 250$~GeV).
	We also considered the situation where masses of 
	all superpartners not exceed 300--400 GeV.
	All corresponding curves together with the absolute
	lower bound from the unconstrained scan are depicted in 
Fig.~\ref{LowBounds}.

	A restriction of the single (light) squark mass to be quite small
	(M$^{}_{\rm sq}< 230$~GeV) as well as another
	assumption that all sferminos masses not exceed 300--400 GeV,
	put upper limits on the mass of the LSP
	and therefore do not permit $R$ to drop very deeply 
	with increasing LSP mass.
	Furthermore  in both cases the lower bound for the
	rate is established for all allowed masses of the LSP
	at a level of $10^{-3}$ events/kg/day.     
	This value for the event rate  we consider as an optimistic 
	sensitivity expectation for high-accuracy future 
	detectors of dark matter, such as GENIUS 
\cite{GENIUS, NECRESST}.
	Practically the same lower bound one obtains under
	the assumption that both charginos are quite light
	(M$^{}_{\rm ch-os}< 250$~GeV).
	In this case LSP masses do not exceed 250 GeV and
	lower rate bound is $2\div5\times 10^{-3}$ events/kg/day.     

\begin{figure}[th!] 
\begin{picture}(100,100)
\put(-10,122){\includegraphics{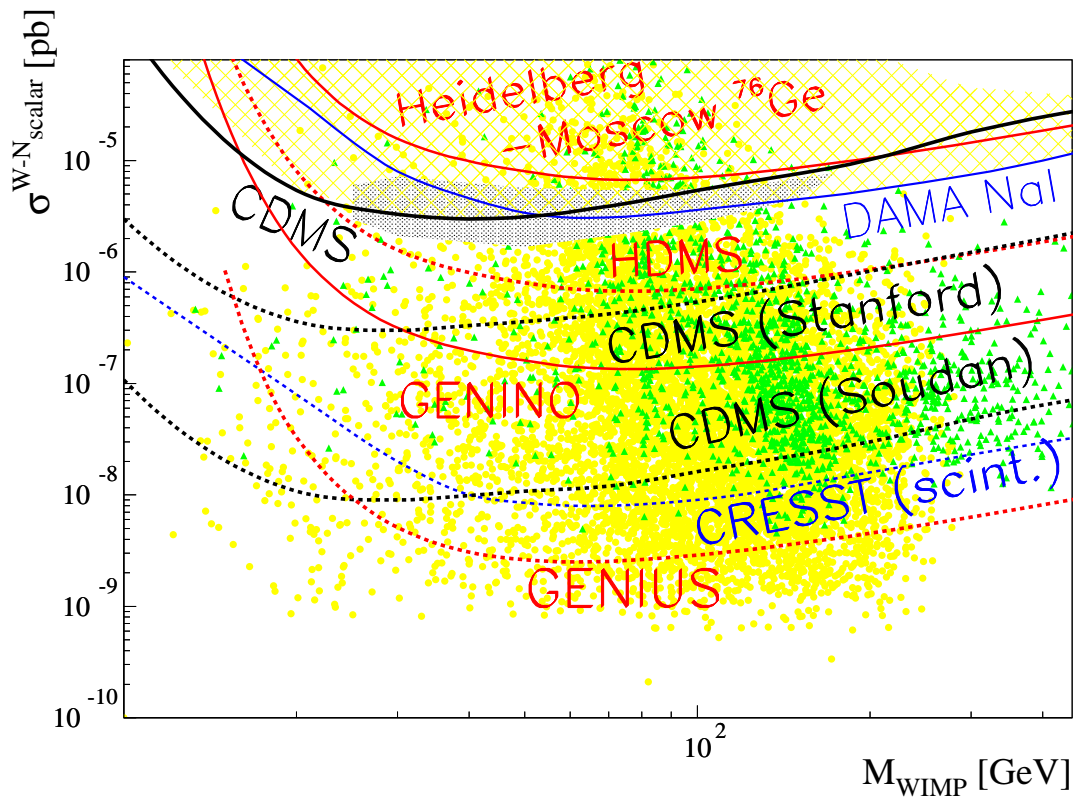}}
\end{picture}
\caption{WIMP-nucleon cross section limits in pb for scalar
	interactions as function of the WIMP mass in GeV.
	Filled circles present our calculations with light SUSY spectrum. 
	Filled triangles give the cross section
	with assumption of M$^{}_{\rm CH}<200$~GeV. 
}
\label{WIMP-nucleon1}
\end{figure}

\begin{figure}[t!] 
\begin{picture}(100,90)
\put(-10,122){\includegraphics{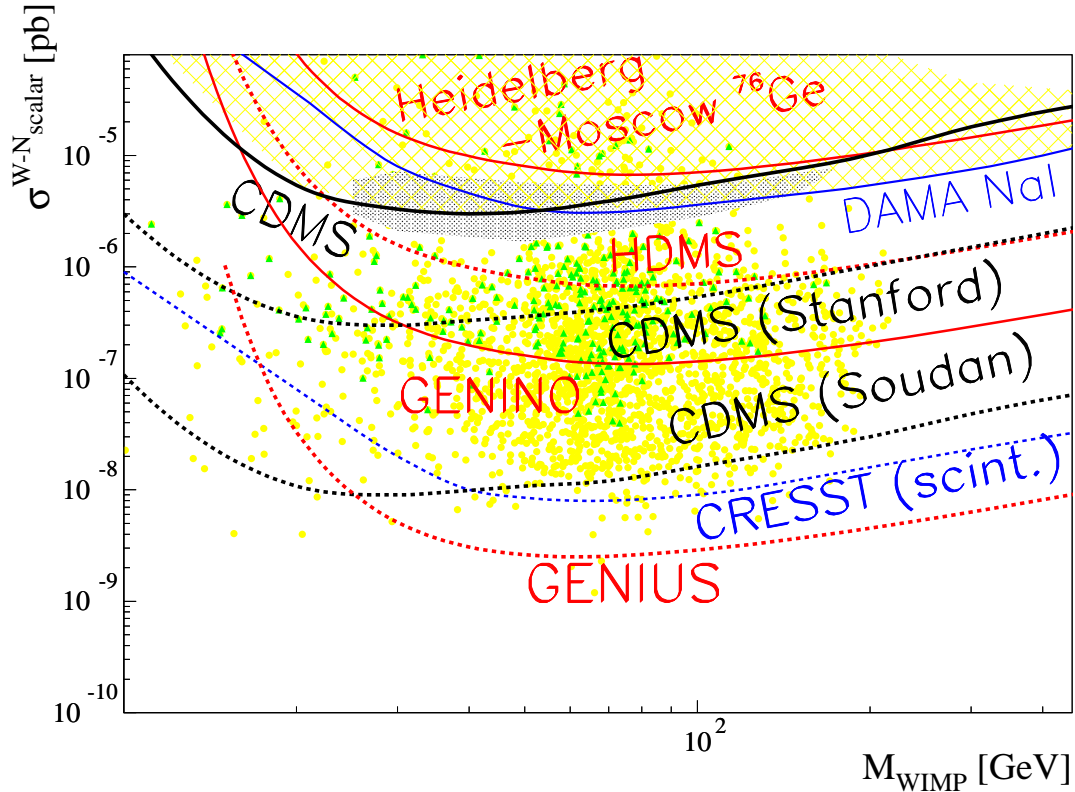}}
\end{picture}
\caption{WIMP-nucleon cross section limits in pb for scalar
	interactions as function of the WIMP mass in GeV.
	Filled circles present our calculations
	with light SUSY spectrum and $\tan\beta>20$. 
	Filled triangles give the same as filled circles, but
	for $\tan\beta>40$. 
}
\label{WIMP-nucleon2}
\vspace*{-10pt}
\end{figure}

\enlargethispage{\baselineskip}
	One can see that the mass of the light neutral CP-even 
	Higgs boson M$^{}_{\rm Hl}$, perhaps most easily measurable 
	experimentally due to its smallness, has unfortunately 
	only a very poor restrictive potential. 
	Already for the mass value M$^{}_{\rm Hl}< 80$~GeV, 
	which is practically equal to the experimental 
	border (M$^{\rm exp}_{\rm Hl}= 79$~GeV),
	the lower bound for the event rate in a detector of 
	$^{73}$Ge is far below $10^{-3}$ events/kg/day.
	The situation looks most promising when one limits the
	mass of the charged Higgs boson.

	From 
Fig.~\ref{Rate-spectrum} one can conclude that in the 
	charged Higgs boson low-mass region other masses of both 
	CP-even and CP-odd Higgs bosons are also restricted from above.
	Therefore coupling constants of the scalar neutralino-quark 
	interaction, which contain terms with 
	$\frac{1}{m^2_{H,h}}$-factors, are not suppressed
	enough and the rate can not decrease significantly.
	The lower bound of the rate increases when the 
	mass M$^{}_{\rm CH}$ decreases 
(Fig.~\ref{Rate-spectrum})
	and for M$^{}_{\rm CH}<200, \ 150$~GeV 	reaches
	values of $\sim 10^{-2}, \ \sim 10^{-1}$ events/kg/day, 
	respectively practically for all allowed masses of the LSP.
	These values can be reached not only with  
	GENIUS ($10^{-2}$ events/kg/day), but also 
	with some other near-future direct dark matter detectors
\cite{HDMS}.
	The fact also is confirmed by 
Fig.~\ref{WIMP-nucleon1}, where 
	cross section limits for the WIMP-nucleon scalar
	interactions for different experiments 
	are presented as functions of the WIMP mass.
	Light-filled circles in 
Fig.~\ref{WIMP-nucleon1} give the scalar cross sections, calculated 
	under the assumption that the SUSY spectrum is quite light
(see Fig~\ref{LowBounds}).
	Filled triangles give the cross section, obtained
	with charged Higgs boson mass restriction M$^{}_{\rm CH}<200$~GeV. 

	Therefore if it happened, for instance,  that 
	either the SUSY spectrum is indeed light, or 
	the charged Higgs boson mass indeed does not exceed 200~GeV,
	in both cases at least 
	the GENIUS experiment should detect a dark matter signal.
	If we consider a more complicated condition,
	for example,  assuming that the SUSY spectrum is quite
	light and simultaneously that $\tan\beta$ is quite large, 
	then not only GENIUS, but also CDMS and HDMS
\cite{NECRESST, HDMS} will possess very good
	prospects to detect a dark matter signal.
	This situation is illustrated in
Fig.~\ref{WIMP-nucleon2}, where besides
	cross section limits for the WIMP-nucleon scalar
	interactions for different experiments 
	are given calculations for the case of a light SUSY spectrum
	with extra assumptions $\tan\beta>20$ (filled circles)
	and $\tan\beta>40$ (filled triangles).

\smallskip
	Therefore the obtained correlations between lower limit
	for the event rate $R$\ and some masses of SUSY particles
	give good prospects for direct dark matter detection
	with next-generation detectors.
	The prospects could be very promising
	if from collider searches one would be able to restrict 
	the mass of the charged Higgs boson at a level of about 200 GeV 
	(light Higgs sector). 
	The observation, due to its importance for 
	dark matter detection, could serve as a source for extra
	efforts in searching for charged Higgs boson with colliders.
	Considered together, both these experiments, 
	collider search for charged Higgs and 
	dark matter search for SUSY LSP,   
	become very decisive for a verification of SUSY models. 

	Otherwise complete non-observation of any dark matter signal
	with very sensitive dark matter detectors in accordance with 
Fig.~\ref{LowBounds}--\ref{WIMP-nucleon2}, would exclude, for example, 
	a SUSY spectrum with masses lighter then 300--400 GeV as well as 
	light SUSY spectrum with large $\tan\beta$
(Fig.~\ref{WIMP-nucleon2}), charginos with masses smaller then 250 GeV 
(Fig.~\ref{LowBounds}), charged Higgs boson with M$^{}_{\rm CH}<200$~GeV,
	and therefore any possibility for the entire light Higgs sector
	in the MSSM
(Fig.~\ref{WIMP-nucleon1}). 	
	
	The last case is in particular interesting, because
	if the light charged Higgs boson is excluded by GENIUS, 
	then either it will be 
	{rather unpromising}
	to search it for with 
	colliders, or any positive result of a collider search
	brings strong contradictions in the MSSM approach 
	to dark matter detections and/or collider SUSY searches.

\bigskip
	Finally we would like to comment the following.
	Unfortunately the MSSM parameter space is huge and to obtain 
	some reliable feeling, concerning, for example, 
	the expected rate of dark matter detection
	when all relevant experimental and cosmological 
	constraints are taken 
	into account, one 
	has nothing but this statistical numerical method
\cite{DressNojiriRate,JuKaGr,9710238,Superlight,Bot}.

 	This method allows lower and upper bounds for
	any observable to be estimated, 
	and to  make conclusions about the prospects 
	for dark matter detection with modern 
	or near-future high-accuracy dark matter detectors. 
	The larger the amount of points which confirms 
	such a conclusion the better. 
	The conclusions we made here are based on hundreds of thousand of
	points which passed all constraints.
	Of course, we have no proved protection against
	peculiar choices of parameters which
	could lead to some cancellation and to 
	small cross sections even if Higgs masses are small. 
	Nevertheless,
	the probability of these choices is very small
	(about 1/100000), otherwise we should already meet them
	with our random scanning. 
	On the  other side, 
	if these peculiar choices exist and one-day would manifest 
	themselves, 
	this would be a very interesting puzzle, 
	because it would be some kind of fine tuning of parameters, 
	which requires strong further development of
	our understanding of the theory.       

 	The results of this paper may be considered as 
	a good example of the complementarity 
	of modern accelerator and non-accelerator 
	experiments looking for new physical phenomena.

\smallskip

	We thank S.G. Kovalenko and L. Baudis for helpful discussions.
	The investigation was supported in part (V.A.B.)
	by Grant GNTP 215 from Russian Ministry of Science.

\clearpage

\end{document}